\documentclass[aps,article,author-year,notitlepage,showpacs]{revtex4-1}
\usepackage{subeqn}
\usepackage{graphicx}
\usepackage{float}
\usepackage[T1]{fontenc}
\usepackage[latin1]{inputenc}
\usepackage{amssymb}
\usepackage{subfigure}
\include{epsf}
\epsfverbosetrue

\newcommand{\bea}{\begin{eqnarray}}
\newcommand{\eea}{\end{eqnarray}}

\makeatother
\begin{document}
\title{Results from the 4PI Effective Action in 2- and 3-dimensions}

\author{M.E. Carrington}
\email[]{carrington@brandonu.ca} \affiliation{Department of Physics, Brandon University, Brandon, Manitoba, R7A 6A9 Canada}\affiliation{Winnipeg Institute for Theoretical Physics, Winnipeg, Manitoba}

\author{Wei-Jie Fu}
\email[]{fuw@brandonu.ca} \affiliation{Department of Physics, Brandon University, Brandon, Manitoba, R7A 6A9 Canada}
\affiliation{Winnipeg Institute for Theoretical Physics, Winnipeg,
Manitoba}

\date{\today}

\begin{abstract}
We consider a symmetric scalar theory with quartic coupling and
solve the equations of motion from the 4PI effective action in 2-
and 3-dimensions using an iterative numerical lattice method. For
coupling less than 10 (in dimensionless units) good convergence is
obtained in less than 10 iterations. We use lattice size up to 16 in
2-dimensions and 10 in 3-dimensions and demonstrate the convergence
of the results with increasing lattice size. The self-consistent
solutions for the 2-point and 4-point functions agree well with the
perturbative ones when the coupling is small and deviate when the
coupling is large.
\end{abstract}


\large
\maketitle

\section{Introduction}
\vspace{5pt}

The resummation of certain classes of Feynman diagrams to infinite loop order
is a powerful method in quantum field theory. A well known example
is the hard thermal loop theory~\cite{Braaten1990}, developed in the context of thermal field
theory, which resums all loop corrections
which are of the same order as tree diagrams, when external momenta are soft.

In recent years, another kind of resummation approach, known as
two-particle irreducible (2PI) effective action theory, has
attracted a lot of attention. In the 2PI formalism, the effective
action is expressed as a functional of the non-perturbative
propagator~\cite{Luttinger1960,Cornwall1974}, which is determined through
a self-consistent stationary equation after the effective action is
expanded to a certain order in the loop or $1/N$ expansion. This self-consistent equation of motion resums certain classes of diagrams to infinite order. The classes that are resummed are determined by the set of skeleton diagrams that are included in the truncated effective action. Numerical studies indicate
that the 2PI effective action formalism is very successful in
describing equilibrium thermodynamics, and also the
quantum dynamics of far from equilibrium of quantum fields.
%Based on the quasiparticle picture,
The entropy of the quark-gluon plasma
obtained from the 2PI formalism shows very good agreement with lattice
data for temperatures above twice the transition
temperature~\cite{Blaizot1999}. The poor convergence problem usually
encountered in high-temperature resummed perturbation theory with
bosonic fields is also solved in the 2PI effective action
theory~\cite{Berges2005a}. Furthermore, it has been shown that
non-equilibrium dynamics with subsequent late-time thermalization can
be well described in the 2PI formalism (see \cite{Berges2001} and references
therein). The 2PI effective action has also been combined with the
exact renormalization group to provide efficient non-perturbative
approximation schemes~\cite{Blaizot2011}. The shear viscosity in the
$O(N)$ model has been computed using the 2PI
formalism~\cite{Aarts2004}.

The 2PI effective action theory has
its own drawbacks and limitations.
When the effective action is
expanded to only 2-loops, the 2PI effective action is complete. However, when the expansion is beyond 2-loops, one must use a higher order effective theory to obtain a complete description \cite{Berges2004}. Higher order effective theories are defined in terms of self-consistently determined $n$-point functions for $n>2$.
It has been shown that the 2PI effective action is insufficient to calculate
transport coefficients for high temperature gauge
theories~\cite{Moore2002,Carrington2006}, but that higher order $n$PI effective actions can be used \cite{Carrington2008}.

The 4PI effective action for scalar field theories is derived in
Ref.~\cite{Norton1975} using Legendre transformations. The method of successive Legendre transforms is used in  \cite{Carrington2004,Berges2004}. A new method has been developed
to calculate the 5-loop 5PI and 6-loop 6PI effective action for
scalar field theories~\cite{Carrington2010,Carrington2011}. The 3PI
and 4PI effective actions have been used to obtain a set of
integral equations from which the leading order and next-to-leading
order contributions to the viscosity can be
calculated~\cite{Carrington2009,Carrington2010b}.

A lot of effort has been devoted to numerical
computations in 2PI effective theories.
%The first step in performing a numerical calculation is the renormalization of the theory. The proof of the renormalizabiltiy of the 2PI effective theory took many years and received contributions from several authors \cite{vanHees2002,Blaizot2003,Serreau2005,Serreau2010}.
For higher order $n$PI theories numerical calculations are extremely difficult and little progress has been made (see however  \cite{Moore2012}).

This paper is organized as follows. In section \ref{generalSection} we define our notation, in section \ref{numericalSect} we present results from our numerical calculations in 2D and 3D, and in section \ref{concSect} we give our conclusions.

\section{General Formalism}
\label{generalSection}
\vspace{5pt}

We consider the following Lagrangian\footnote{The coupling constant $\lambda$ is imaginary. Using this definition the lines and crosses in Feynman diagrams are propagators and proper vertex functions, as defined in equation (6), and the diagrams in Figures 1-3 do not carry signs or extra factors of $i$. Numerical calculations are done in Euclidean space and the corresponding vertex is defined in equation (\ref{euc}).}
\begin{equation}
{\mathcal{L}}=\frac{1}{2}(\partial_{\mu}\varphi\partial^{\mu}\varphi-m^{2}{\varphi}^{2})
-\frac{i\lambda}{4!}\varphi^{4}\,.
\end{equation}
The classical action is:
\bea
\label{defnClass}
&& S[\phi]=S_{0}[\varphi]+S_{\mathrm{int}}[\varphi]\,,\\
&& S_{0}[\varphi]=\frac{1}{2}\int d^d x\, d^dy\,\varphi(x)\big[i G_0^{-1}(x-y)\big]\varphi(y) \,,\nonumber\\
&& S_{\mathrm{int}}[\varphi]=- \frac{i\lambda}{4!}\int d^d x  \varphi^4 (x)\,.\nonumber
\eea
In most equations in this paper, we suppress the arguments that denote the space-time dependence of functions. As an example of this notation, the non-interacting part of the classical action is written:
\bea
\label{notex}
&& S_{0}[\varphi]=\frac{1}{2}\int d^d x\, d^dy\,\varphi(x)\big[i G_0^{-1}(x-y)\big]\varphi(y)~~ \rightarrow ~~ \frac{i}{2}G_0^{-1}\varphi^2\,,\\
&&G_0^{-1}=-i \frac{\delta^2 S_{cl}[\varphi]}{\delta \varphi^2}\bigg|_{\varphi=0}=i(\Box+m^2)\,.\nonumber
\eea

The effective action is obtained from the Legendre transformation of the connected generating functional:
\bea
\label{genericGamma}
&& Z[R_1,R_2,R_3,R_4]=\int [d\varphi]  \;{\rm Exp}[i\,(S_{cl}[\varphi]+R_1\varphi + \frac{1}{2} R_2\varphi^2 + \frac{1}{3!} R_3\varphi^3 +\frac{1}{4!} R_4\varphi^4)]\,,\\[1mm]
&&W[R_1,R_2,R_3,R_4]=-i \,{\rm Ln} Z[R_1,R_2,R_3,R_4]\,,\nonumber\\[1mm]
&&\Gamma[\phi,G,V_3,V_4] = W - R_1\frac{\delta W}{\delta R_1} - R_2\frac{\delta W}{\delta R_2} - R_3\frac{\delta W}{\delta R_3} - R_4\frac{\delta W}{\delta R_4} \,.\nonumber
\eea
Connected and proper Green functions are denoted $V_j^c$ and $V_j$ respectively, where the subscript $j$ indicates the number of legs. They are defined:
\bea
\label{Wders}
&& V_j^c = \langle\varphi^j\rangle_c = -(-i)^{j+1}\frac{ \delta^j W}{\delta R_1^j}\,,\\[2mm]
&& V_j= i \frac{\delta^j}{\delta \phi^j} \Gamma_{\rm 1PI} = i \frac{\delta^j}{\delta \phi^j}\left(W[R_1]-R_1 \phi\right)\,.
\eea
The equations that relate the connected and proper vertices are obtained from their definitions using the chain rule.
We organize the calculation of the effective action using the
method of subsequent Legendre transforms \cite{Carrington2004,Berges2004}. This method involves starting from an expression for the  2PI effective action and exploiting the fact that the source terms $R_3$ and $R_4$ can be combined with the corresponding bare vertex by defining a modified interaction vertex. The result is:
\bea
&& \Gamma[\phi,G,V_3,V_4]= \frac{i}{2}G_0^{-1}\phi^2 + \frac{i}{2}\mathrm{Tr}\ln
G^{-1}+\frac{i}{2}\mathrm{Tr}G^{-1}_{0}G+\Gamma_{\mathrm{int}}[\phi,G,V_3,V_4]\,,\label{Eq4} \\
&& i\Gamma_{\rm int}[\phi,G,V_3,V_4] =  \frac{\lambda}{4!}\phi^4 +
\frac{\lambda}{4}(\phi^2 G)+\Phi_2[\phi,G,V_3,V_4]\,,\nonumber
\eea
where $\Phi_2[\phi,G,V_3,V_4]$ represents diagrams with 2 and more loops.
In this paper we consider only the self-consistent 2- and 4-point functions in the symmetric phase. These are obtained by solving simultaneously the equations of motion:
\bea
\label{equation:eom}
\frac{\delta \Gamma[\phi,G,V_3,V_4]}{\delta G}\bigg|_{\phi=0,G=\tilde G,V_3=0,V_4=\tilde V_4}=0\,,~~~\frac{\delta \Gamma[\phi,G,V_3,V_4]}{\delta V_4}\bigg|_{\phi=0,G=\tilde G,V_3=0,V_4=\tilde V_4}=0\,.
\eea
From now on we drop the subscript on the 4-point vertex function and write $V:=V_4$ and $\tilde V:=\tilde V_4$.

\vspace*{2mm}
%\section{Renormalization}
%\label{renormSection}

The variables $m$, $\lambda$, $G$ and $V$ in section \ref{generalSection} should all carry a subscript $B$ to indicate that they are bare quantities.
These bare quantities (with subscript $B$) are related to the
renormalized ones by the following relations:
\begin{eqnarray}
\label{Zdefn}
&& \delta m^{2} = Zm_{B}^{2}-m^{2}\,,~~
\delta \lambda = Z^{2}\lambda_{B}-\lambda\,,\\[2mm]
&& G_{B} = ZG\,,~~V_{B} = Z^{-2}V \,,\nonumber \\[2mm]
&& ZG^{-1}_{0B} =  G^{-1}_{0}+ \delta G^{-1}_{0}\,,~~~\delta G^{-1}_{0}=i(\delta Z\Box+\delta m^2)\,,~~~\delta Z = Z-1\,.\nonumber
\end{eqnarray}
In order to simplify the notation we have not introduced a subscript $R$ for renormalized quantities and we have suppressed the subscript $B$ everywhere except in the equation (\ref{Zdefn}). All quantities in section \ref{generalSection} are bare, and all quantities in the following sections are renormalized.

\begin{figure}[H]
\begin{center}
\includegraphics[width=16cm]{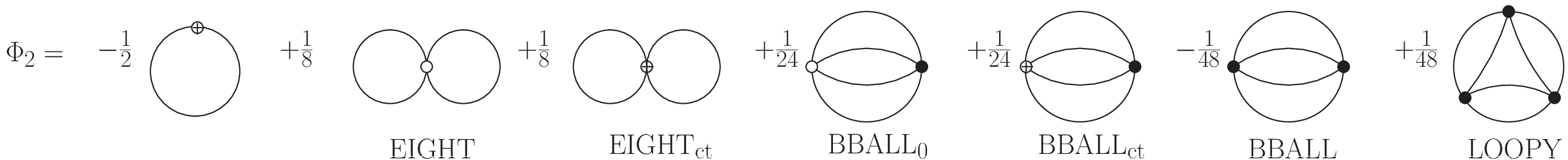}
\end{center}
\caption{The effective action at 4-Loops including counter-terms.
The open circle is a bare vertex, the circle with a cross denotes a
counter-term, and the vertex $V$ is indicated by a solid dot.
\label{PhiAandB}  }
\end{figure}

We divide the functional $\Phi_2[G,V]$ into two pieces: terms without counter-terms or bare vertices, and terms that do contain either counter-terms or bare vertices. We denote these two pieces $-i\Phi_{\mathrm{int}}[G,V]$ and $-i\Phi_0[G,V]$, respectively. Using this notation we write the effective action in equation (\ref{Eq4}) as:
\bea
\label{gammaDef}
 i\Gamma[G,V]&&=- \frac{1}{2}\mathrm{Tr}\ln G^{-1} - \frac{1}{2}\mathrm{Tr}G^{-1}_{0} G + \Phi_0[G,V] + \Phi_{\mathrm{int}}[G,V]\,,\\
\Phi_0 &&= - \frac{1}{2}\mathrm{Tr}\delta
G^{-1}_{0}G+ \frac{1}{8}(\lambda+\delta\lambda) G^2  + \frac{1}{4!}(\lambda+\delta\lambda) G^4 V\,.\nonumber
\eea
The functional $\Phi_0$ has the same form for $n\ge 4$ and all orders in the loop expansion and $\Phi_{\mathrm{int}}$ contains a set of skeleton diagrams which are determined by the orders of the Legendre transform and loop expansion. The sum $\Phi_2=\Phi_0+\Phi_{\mathrm{int}}$ is shown to 4-loops in figure \ref{PhiAandB}.\footnote{All figures are drawn using jaxodraw \cite{jaxo}.} In all diagrams, bare 4-vertices
are represented by white circles, counter-terms are circles with
crosses in them, and solid dots are  the vertex $V$.
\par

\begin{figure}[H]
\begin{center}
\includegraphics[width=14cm]{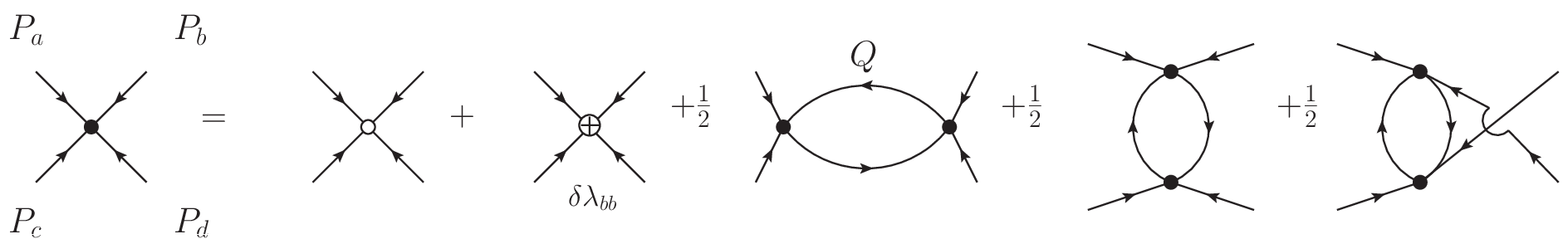}
\end{center}
\caption{Self-consistent equation of motion for the proper
4-point vertex.\label{f4}}
\end{figure}

The equation of motion for $V$ is obtained from the variational equation $\delta\Gamma/\delta V=0$ (see equation (\ref{equation:eom})). Using the 4-Loop 4PI effective action gives the result in figure \ref{f4} and equation (\ref{Eq15}). We use throughout the notation $dQ=d^d q/(2\pi)^d$.
%In order to shorten some of the figures and formulas which appear in this paper, we sometimes combine contributions which correspond to different permutations of external legs. We use bracketed numerical factors to indicate the presence of permutations that are not explicitly written. In figure \ref{f44} we illustrate this notation: the three 1-loop diagrams in figure \ref{f4} are combined.
\par
\bea
V(P_{a},P_{b},P_{c})&& =\lambda+\delta\lambda+V_{s}(P_{a},P_{b},P_{c})
+V_{t}(P_{a},P_{b},P_{c})+V_{u}(P_{a},P_{b},P_{c}) \label{Eq15}\,,\nonumber\\[4mm]
V_{s}(P_{a},P_{b},P_{c})&&=\frac{1}{2}\int dQ\; V(P_{a},P_{c},Q)G(Q)G(Q+P_{a}+P_{c})V(P_{b},P_{d},-Q)\,, \\
V_{t}(P_{a},P_{b},P_{c})&&=V_{s}(P_{a},P_{c},P_{b})\,,\nonumber\\[2mm]
V_{u}(P_{a},P_{b},P_{c})&&=V_{s}(P_{a},P_{b},P_{d})\,,~~~P_{d}=-(P_{a}+P_{b}+P_{c})\,.\nonumber
\eea
% and $dQ$ is assumed to represent $d^{d}Q/(2\pi)^{d}$ in $d$-dimensions.

\begin{figure}[H]
\begin{center}
\includegraphics[width=16cm]{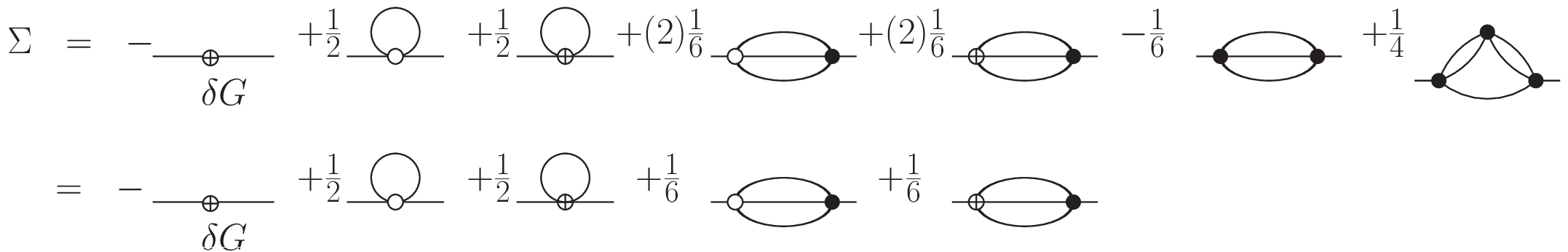}
\end{center}
\caption{The self-energy obtained from equations (\ref{gammaDef}) and (\ref{G2defn}).}\label{SEall}
\end{figure}

The equation of motion for the 2-point function is obtained from the variational equation $\delta\Gamma/\delta G=0$ (see equation (\ref{equation:eom})). It has the form of a Dyson
equation where the self-energy is proportional to the functional derivative of the terms in the effective action with two and more loops: \bea
\label{G2defn} &&G^{-1}=G_0^{-1}-\Sigma\,,~~\Sigma =
2\frac{\delta \Phi_2}{\delta G} \,. \eea
The result is shown in the first line of figure \ref{SEall}. The diagrams can be rearranged by substituting the $V$ equation of motion into the vertex on the left side of the sixth diagram. %This produces two terms, one of which is a sunset diagram with two proper vertices, and the other cancels the 2-loop skeleton diagram in figure \ref{SEall}.
This substitution cancels the 3-loop diagram and produces the result in the second line of the figure and equation (\ref{Eq19}).
\par
\begin{eqnarray}
\Sigma(p) &=&i(\delta Z P^{2}-\delta
m^{2})+\frac{1}{2}(\lambda+\delta\lambda)\int dQ\; G(Q)\nonumber \\
&&+\,\frac{1}{6}(\lambda+\delta\lambda)\int dQ \int dK\;
V(P,Q,K)G(Q)G(K)G(Q+K+P)\,.\label{Eq19}
\end{eqnarray}

\vspace*{2mm}

Numerical calculations will be done in Euclidean space and therefore we redefine the variables:
\bea 
\label{euc}
&& q_0=i q_E\,,~~ \delta Z = \delta Z_E\,,~~\delta m^2=\delta m^2_E\,,~~
G=-iG_{E}\,,~~\Sigma=-i\Sigma_{E}\,,\\
&& \lambda = -i\lambda_E\,,~~\delta\lambda=
-i\delta\lambda_E\,,~~V=iV_{E}\,.\nonumber
\eea
The Dyson equation in Euclidean space is (see equation (\ref{G2defn})):
\begin{equation}
G_E^{-1}(P)=G^{-1}_{0E}(P)+\Sigma_E(P)\,.\label{Eq88}
\end{equation}
All variables from here on are Eucledian and we suppress the subscript $E$.
The coupling constant and 4-point vertex have dimension
$4-d$ so that in 2D $\lambda\sim m^2$ and in 3D $\lambda\sim m$. We work in mass units, in which all dimensional quantities are scaled by the mass.

\section{Numerical Results}
\label{numericalSect}

\subsection{Perturbative Theory}

We start by looking at the perturbative theory.  For the 4-point function at the 1-loop level the diagrams we need are obtained from figure \ref{f4} with lines and proper vertices replaced by bare ones. In less than 4-dimensions there are no
ultraviolet divergences and from equation (\ref{Eq15}) we find:
\begin{eqnarray}
\label{vpert}
V(P_{a},P_{b},P_{c})&=&-\lambda-\delta\lambda +\frac{\lambda^{2}}{2}\frac{\Gamma(2-d/2)}{(4\pi)^{d/2}}\int_{0}^{1}d
x
\bigg\{\frac{1}{\big[m^{2}+x(1-x)(P_{a}+P_{c})^{2}\big]^{2-d/2}}\nonumber \\
&&+\frac{1}{\big[m^{2}+x(1-x)(P_{a}+P_{b})^{2}\big]^{2-d/2}}+\frac{1}{\big[m^{2}+x(1-x)(P_{b}+P_{c})^{2}\big]^{2-d/2}}\bigg\}\,.
\end{eqnarray}

% and the
%counter-term $\delta \lambda$ determined by the condition (\ref{renormc3}) merely shifts the finite part so that the zero %XXmomentum vertex is normalized to the bare value.

To obtain the 2-point function at 2-Loops we need to calculate the tadpole and sunset diagrams in equation (\ref{Eq19}) with lines and proper vertices replaced by bare ones.
%The diagrams we need for the 2-point function are shown in figure \ref{f5}.
%\par\begin{figure}[H]
%\begin{center}
%\includegraphics[width=10cm]{Sigma1.eps}
%\end{center}
%\caption{1-loop and 2-loop self-energy diagrams in perturbation
%theory.\label{f5}}
%\end{figure}
We use dimensional regularization and define $\epsilon=1-d/2$. The tadpole diagram is momentum independent in any number of dimensions. In 3D it is finite using dimensional regularization and in 2D the divergent part is easily obtained as $\lambda/(8\pi \epsilon)$. The sunset contribution to the self-energy is:
\begin{eqnarray}
\label{spert}
&&\Sigma^{\rm 2~dims}_{\rm sunset}(P) = -\frac{\lambda^{2}}{6(4\pi)^{2}}\int_{0}^{1}dx
\int_{0}^{1}dy\frac{1}{[y+(1-y)x(1-x)]m^{2}+y(1-y)x(1-x)P^{2}}\,.\label{wj1}\\
&&\Sigma^{\rm 3~dims}_{\rm sunset}(P)=-\frac{\lambda^2}{6}\frac{\Gamma(\epsilon)}{(4\pi)^{3-2\epsilon}}\int_0^1 \frac{dx}{(x(1-x))^{1/2+\epsilon}}\int_0^1 \frac{dy\,y^{\epsilon-1/2}}{[P^2y(1-y)+m^2(1-y+\frac{y}{x(1-x)})]^{\epsilon}}\,.\nonumber
\end{eqnarray}
The integral is finite in 2D. In 3D the divergence is momentum independent. We can write the divergent part of the self-energy as:
\bea
\label{m-2loop}
\Sigma^{\rm div} = \frac{\lambda}{8\pi}\,\frac{1}{\epsilon}\delta_{d2} -\frac{1}{\epsilon}\frac{\lambda^2}{12(4\pi)^2}\delta_{d3}\,,
\eea
 where the Kronecker deltas indicate which pieces contribute in 2D and 3D.
Thus we have that in 2D the tadpole has a momentum independent divergence and the sunset diagram is finite, while in 3D the situation is reversed and the sunset has a momentum independent divergence but the tadpole is finite.
Using Pauli-Villars regularization the tadpole has a momentum independent divergence in 2D or 3D and the sunset has a momentum  independent divergence in 3D only. In all cases, the counter-term $\delta m^2$ completely removes the divergence and we can set $\delta Z=0$.
We use the renormalization
condition $
\Sigma(0)=0$ to determine the mass counter-term.
%We use the condition $\Sigma(0)=0$ to determine the mass counter-term.
Since the tadpole diagram is independent of the
external momentum (in any dimension), this renormalization condition completely
removes the entire tadpole contribution, and we can just drop the
diagram.
In both 2D and 3D the 4-vertex does not UV-renormalize and the natural choice is to set $\delta\lambda=0$, so that $\lambda$ is defined as the limit of the 4-point function at large external momenta. 

\subsection{Non-perturbative calculation}

%Expanding the coupled set of integral equations in figures \ref{f4} and \ref{SEall} produces an infinite set of diagrams.
The diagrams produced by expanding the $n$PI equation of motion
are not the same as those produced by the perturbative expansion, some diagrams appear with different symmetry factors, and some diagrams are missing altogether. In less than 4-dimensions however, the only fundamental divergences are the tadpole and sunset diagrams, and each insertion of a bare self-energy is accompanied by the mass counter-term that makes it finite. Iteration does not create new sub-divergences and therefore iterations amount to inserting renormalized self-energies, without introducing new divergences. Therefore one can also renormalize the non-perturbative theory using only a mass counter-term.
In order to compare the non-perturbative results with the perturbative ones, we use the same renormalization conditions.

To obtain non-perturbative results we solve the self-consistent
equation of motion for the 2- and 4-point functions using a
numerical lattice method. We use an $N^{d}$
symmetric lattice with periodic boundary conditions. The size of the lattice is limited by the calculation time and memory constraints.
In 2D we use $N$ up to 16 and in 3D $N$ up to 10.
The lattice
spacing is $a$ and we choose $a=2\pi/(N m)$.
In
Euclidean space, each momentum component is discretized:
\begin{equation}
p_{i}=\frac{2\pi}{aN}n_{i}\,,\quad
n_{i}=-\frac{N}{2}+1,...,\frac{N}{2}\,,\label{momenta}
\end{equation}
and the periodic boundary conditions take the form $n_i+N=n_i$ for all momentum components. 
%We emphasize that in our calculations we use a momentumlattice and not a spatial one.
The lattice momenta given by equation~(\ref{momenta}) form a
Brillouin zone. On the lattice, the equation of motion for the
4-point vertex (equation (\ref{Eq15})) is
transformed into:
\begin{eqnarray}
V(P_{a},P_{b},P_{c})&=&-\lambda - \delta\lambda +\frac{1}{2}\frac{1}{(aN)^{d}}\sum_{Q}\Big[V(P_{a},P_{c},Q)G(Q)G(Q+P_{a}+P_{c})V(P_{b},P_{d},-Q)\nonumber \\
&&+V(P_{a},P_{b},Q)G(Q)G(Q+P_{a}+P_{b})V(P_{c},P_{d},-Q)\nonumber \\
&&+V(P_{a},P_{d},Q)G(Q)G(Q+P_{a}+P_{d})V(P_{b},P_{c},-Q)\Big]\,,\label{Eq25}
\end{eqnarray}
and the 2-loop self-energy (equation (\ref{Eq19})) is:
\begin{eqnarray}
\Sigma(P) &=&\delta Z \,P^2+ \delta m^2 + \frac{1}{6}(\lambda+\delta\lambda)
\frac{1}{(aN)^{2d}}\sum_{Q}\sum_{K}V(P,Q,K)G(Q)G(K)G(Q+K+P)\,.\label{Eq26}
\end{eqnarray}
We start from an initial 4-point
vertex and propagator which we chose to be the bare
vertex and free propagator. Then we use equations (\ref{Eq88}),
(\ref{Eq25}) and (\ref{Eq26}) and  simultaneously search
for self-consistent solutions using an iterative procedure. In order to
make the iterations converge more quickly, we adopt the following
formula to update the vertex and propagator at every
iteration~\cite{Berges2005a}:
\begin{eqnarray}
V_{\mathrm{update}}&=&\alpha V_{\mathrm{new}}+(1-\alpha)V\,, \\
G_{\mathrm{update}}&=&\alpha G_{\mathrm{new}}+(1-\alpha)G\,,\label{}
\end{eqnarray}
where $\alpha$ is the convergence factor, and we choose
$\alpha=0.8$. 
%added

In all of our calculations, the full momentum dependence of the vertex and self-energy is taken into account. 
In order to produce figures, we must fix some momentum components to obtain a 2-dimensional representation of the results. For the 4-point function, when we consider the dependence on either the number of iterations or the coupling, we choose all momentum components equal to zero. We also study the momentum dependence of the 4-point function as a function of $(p_a)_x$ and $\{(p_a)_x,(p_b)_x\}$ with all other momentum components set to zero, where $(p_a)_x$  and $(p_b)_x$  are the $x$-components of the momentum of the first and second legs. For the 2-point function, because of the renormalization condition, we consider $\Sigma(p_x=2,0,0)$ as a function of the number of iterations and coupling, and also $\Sigma(p_x,0,0)$ at fixed coupling.

\begin{figure}[H]
\begin{center}
\includegraphics[scale=1.2]{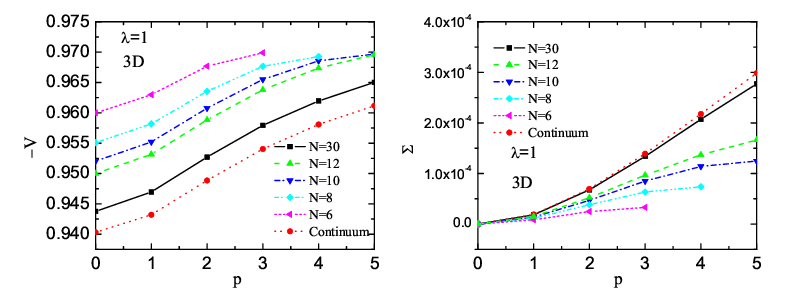}
\caption{The perturbative vertex and self-energy from the lattice calculation with $N=6$, 8, 10, 12 and 30, and in the continuum limit. }\label{fig:Pert3D}
\end{center}
\end{figure}

It is interesting to compare the results we obtain from the non-perturbative calculation with the corresponding perturbative ones.
The continuum perturbative solution can be obtained from equations (\ref{vpert}) and (\ref{spert}).
In order to check our equations, we also do the perturbative calculation on the lattice by solving equations (\ref{Eq25}) and (\ref{Eq26}) with the the self-consistent vertex and propagator replaced by the bare ones. 
We work in 3D and use $N=6$, 8, 10, 12 and 30, the results are shown in figure \ref{fig:Pert3D}. For very large $N$ the lattice calculation converges to the continuum limit, as it should. In this paper we do not go beyond $N=10$, and although it is clear that larger $N$'s are desirable, the figure shows that the calculation converges in the sense that $N=10$ is closer to $N=8$ than $N=8$ is to $N=6$. Later in this section we discuss convergence further. 

\begin{figure}[H]
\begin{center}
\includegraphics[scale=1.2]{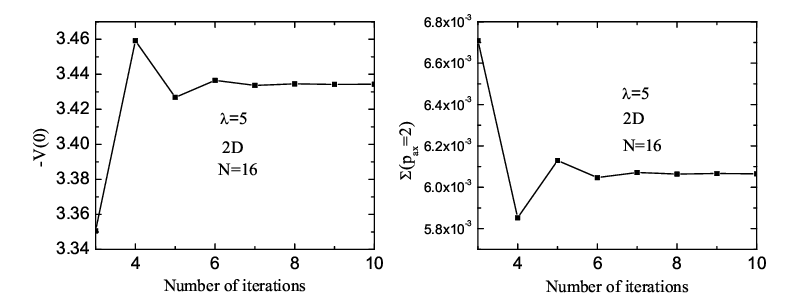}
\includegraphics[scale=1.2]{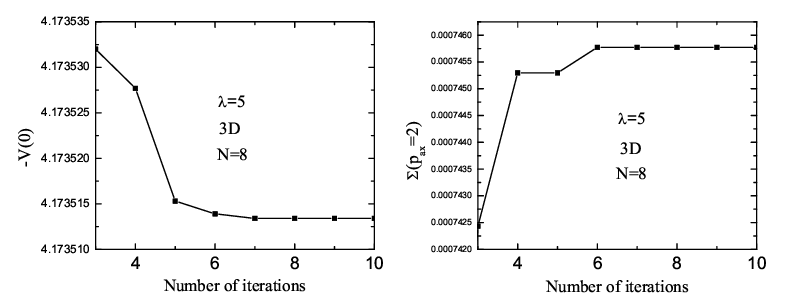}
\caption{The 4-point vertex and self-energy as a function of the
number of iterations for $\lambda=5$, and $N=16$ for 2D and $N=8$
for 3D.}\label{fig:conver23D}
\end{center}
\end{figure}

In figure \ref{fig:conver23D} we show the 4-point function and self-energy as a function of the number of iterations.
We choose $\lambda=5$ (in mass units), and $N=16$ in 2D and $N=8$ in 3D. The first two iterations are not included so that the evolution can be seen more clearly. 
In both the 2D and 3D cases, self-consistent
convergent solutions are obtained quickly after several iterations.
The number of iterations that is needed increases
slightly as $\lambda$ increases, but it is easy to obtain converged
solutions for $\lambda \sim 10$.

\begin{figure}[H]
\begin{center}
\includegraphics[scale=1.3]{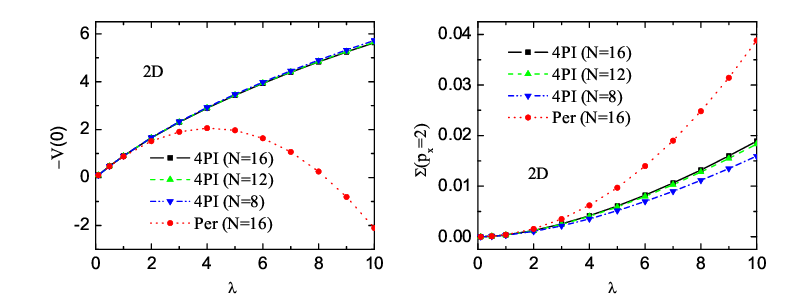}
\includegraphics[scale=1.3]{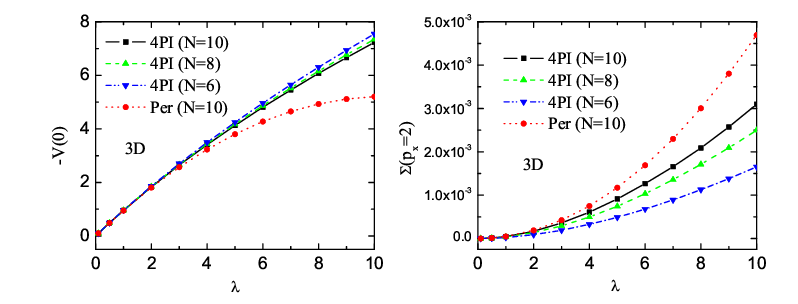}
\caption{Comparison of the 2D and 3D 4-point vertex and self-energy
as functions of the coupling strength $\lambda$. The 4PI
calculations are done in 2D with $N=16,12$ and 8, and in 3D with
$N=10,8$ and 6. 
%We use $(p_a)_x = 2$ and all other momenta components zero. 
The top left panel is the 2D 4-vertex, top right is
the 2D self-energy,  bottom left is the 3D 4-vertex, bottom right is
the 3D self-energy. In each graph the perturbative result is the
dotted line which joins the round markers (red on-line). }\label{f6}
\end{center}
\end{figure}

Figure \ref{f6} shows the 2D and 3D 4-point vertex and 
self-energy as functions of the coupling strength $\lambda$,
calculated from 4PI effective action and perturbation theory. 
In all cases the non-perturbative results agree well with the perturbative ones when the coupling strength is small. 
When the coupling
constant becomes large the 4PI results differ significantly
from the perturbative ones, indicating the importance of a non-perturbative method in the strong coupling regime. 
%Non-perturbative results are shown in 2D with
%$N=16,12$ and 8, and in 3D with  $N=10, 8$ and $6$. 
%Convergence is better for the vertex than for the self-energy, which is expected since for the 4-point function in less than 4-dimensions all integrals converge in the ultra-violet. 
In 2D, the results for the 4PI vertex are almost
independent of the lattice number $N$. The self-energy depends more strongly on the lattice size but converges well when $N$ is increased to 16 (the curve corresponding to $N=16$ almost coincides
with that corresponding to $N=12$).
%, as the top right panel of Fig.
%\ref{f6} shows. Therefore, we believe that our non-perturbative
%results are robust and reasonable. 
In 3D, calculations can only be performed up to
$N=10$ because the three
independent external momenta of the 4-point vertex consume lot of
computational resource. Convergence is good for the 4-point vertex, but the results for the self energy have a stronger dependence on the lattice
number.

\begin{figure}[H]
\begin{center}
\includegraphics[scale=1.2]{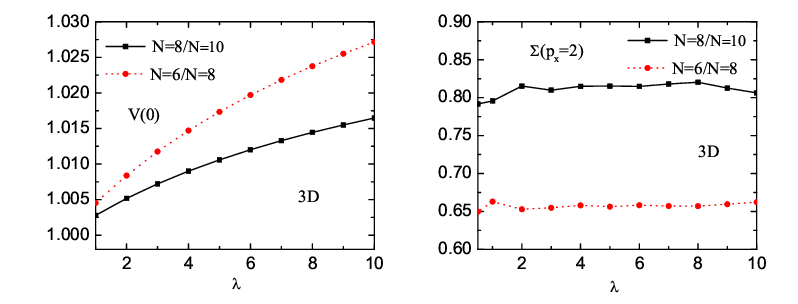}
\caption{Non-perturbative results for 
$V_{N=8}/V_{N=10}$, $V_{N=6}/V_{N=8}$, $\Sigma_{N=8}/\Sigma_{N=10}$ and $\Sigma_{N=6}/\Sigma_{N=8}$ in 3D
as functions of the coupling strength $\lambda$. }\label{fig:ratio}
\end{center}
\end{figure}

In order to investigate whether 
convergence is obtained with $N=10$, we look at the ratios of the vertex and self
energy calculated with $N=8$ and $N=10$, and with
$N=6$ and $N=8$.  The results are shown in Fig. \ref{fig:ratio}. 
%For the self-energy, these ratios are almost independent of the coupling strength and of order $\Sigma_{N=6}/\Sigma_{N=8} \sim 0.65$ and $\Sigma_{N=8}/\Sigma_{N=10}\sim 0.8$ . 
The fact that these ratios approach 1 indicates that our results are converging. 
%This result
%indicates that we indeed approach to convergence with the increase
%of the lattice number and $N=10$ is almost enough to get the
%convergent results since $\Sigma_{N=8}/\Sigma_{N=10}$ is about 0.8
%and close to 1 and we have good reason to believe that the next
%ratio, for example $\Sigma_{N=10}/\Sigma_{N=12}$ to be closer to 1
%comparing with $\Sigma_{N=8}/\Sigma_{N=10}$. In the same way, one
%sees that in 3D the non-perturbative results and the perturbative
%ones match very well at small coupling while develop difference at
%large $\lambda$, especial for the 4-point vertex.

\begin{figure}[H]
\includegraphics[scale=1.3]{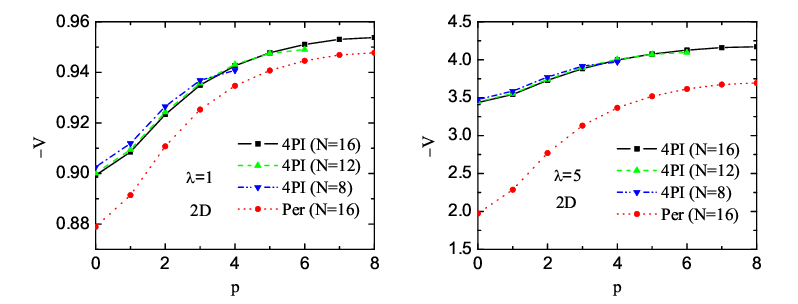}
\includegraphics[scale=1.3]{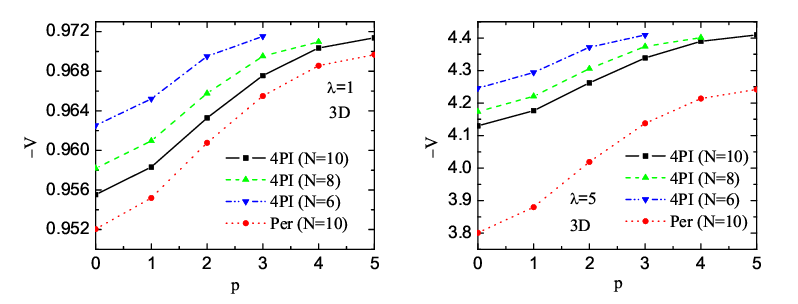}
\caption{Dependence of the 2D and 3D 4-point vertex  on $(p_a)_x$
with all other external momenta components set to zero. Top left
panel is 2D with $\lambda=1$, top right is 2D with $\lambda=5$,
bottom left is 3D with $\lambda=1$, and bottom right is 3D with
$\lambda=5$. The 4PI calculations are done in 2D with $N=16,12$ 8,
and in 3D with $N=10,8$ and 6. 
The perturbative result is the
dotted line which joins the round markers (red on-line).}\label{f7}
\end{figure}

Figure \ref{f7} shows the dependence of the 4-point vertex on the
first momentum component in 2D and 3D, choosing all momentum
components other than $(p_a)_x$ zero and using the coupling
strength $\lambda$ equal to 1 and 5.  We
choose $N=16,12$ and 8 for the 2D calculations, and $N=10,8$ and 6
for the 3D ones. 
The difference between the non-perturbative vertex and the perturbative one is greater when the coupling constant is larger, as expected.The momentum dependence is produced by the 1-loop diagram and at large momentum the 4-point vertex scales as $-V\sim \lambda - C p^{d-4}$, as expected. 
% At large momenta all results agree due to the renormalization conditions.
The results obtained by increasing the lattice number converge in both 2D
and 3D.

\begin{figure}[H]
\includegraphics[scale=1.3]{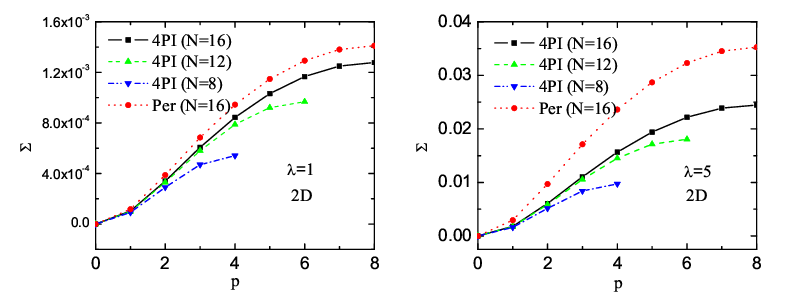}
\includegraphics[scale=1.3]{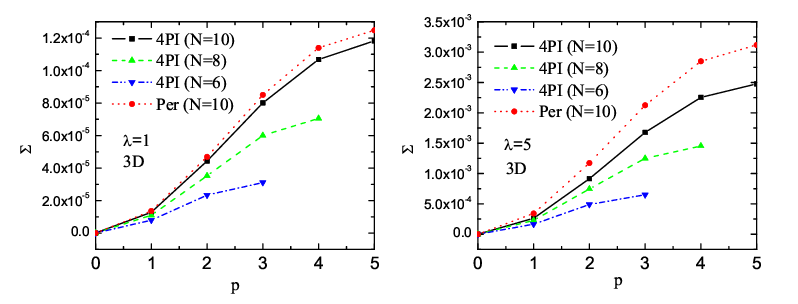}
\caption{Dependence of the 2D and 3D self-energy on the $p_x$ with
all other momentum components set to zero. Top left panel is 2D with
$\lambda=1$, top right is 2D with $\lambda=5$, bottom left is 3D
with $\lambda=1$, and bottom right is 3D with $\lambda=5$.  The 4PI
calculations are done in 2D with $N=16,12$ 8, and in 3D with
$N=10,8$ and 6. 
The perturbative result is the
dotted line which joins the round markers (red on-line).}\label{f8}
\end{figure}

In figure \ref{f8}  we show the dependence of the self-energy on 
$p_x$ with all other momentum components zero. The momentum dependence comes from the sunset diagram and at large momentum the self-energy scales like $C-C^\prime/p^2$ in 2D and $C+C^\prime{\rm log}(p)$ in 3D, as expected. 
The difference between the non-perturbative self energy and the perturbative one is greater when the coupling constant is larger.
Convergence with increasing lattice size is not as good as for the vertex, and not as good in 3D as in 2D. However, the analysis in figure \ref{fig:ratio} indicates that our results are converging. 

\begin{figure}[H]
\begin{center}
\includegraphics[width=14cm]{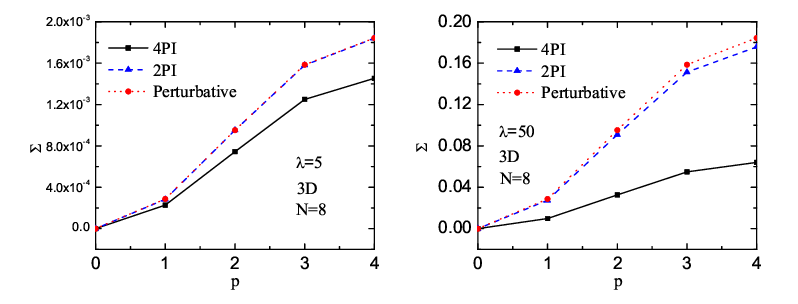}
\end{center}
\caption{Comparison of the 3D perturbative, 2PI and 4PI self-energies for $\lambda=5$ and $\lambda=50$, with $N=8$.\label{fig:Sig_com}}
\end{figure}

It is interesting to compare the 4PI 2-point function with the 2PI version, which is obtained from equation (\ref{Eq26}) with the self-consistent vertex replaced by the bare one. For the values of $\lambda$ chosen in figure \ref{f8} the 2PI result is almost identical to the perturbative one, and it is only for very large values of $\lambda$ that one can see the difference. We illustrate this point in figure \ref{fig:Sig_com} where we show the perturbative, 2PI and 4PI self-energies for $\lambda=5$ and 50, and $N=8$. 
\par

%In 2D the self-energy
%calculated from the 4PI theory is consistent with the perturbative one
%when the coupling strength is small, {\it and it is suppressed as
%$P^{-2}$ at large momentum - i don't see this.} Results 
% converge well with increasing lattice number except for
%the values near the boundary momenta. 
%, which indicates that the lattice number $N=16$ is large enough to obtain convergent results.
 %in 2D
%deviates from the perturbative one obviously at large $\lambda$,
%however it is still suppressed at large momentum. 

%In 3D the self
%energy is expected to be $\sim \ln P$ at large momentum, which is
%also observed in the non-perturbative results away from the boundary
%momenta. We have discussed above that 4PI calculations in 3D with
%lattice number $N=10$ are close to the convergent ones, especially
%when results are not near the boundary momenta, and these results
%deserve to be believed. However, in order to obtain definite results
%at large momenta, it is required to extend the 4PI calculations in
%3D to larger lattice number.

\begin{figure}[H]
\includegraphics[scale=0.7]{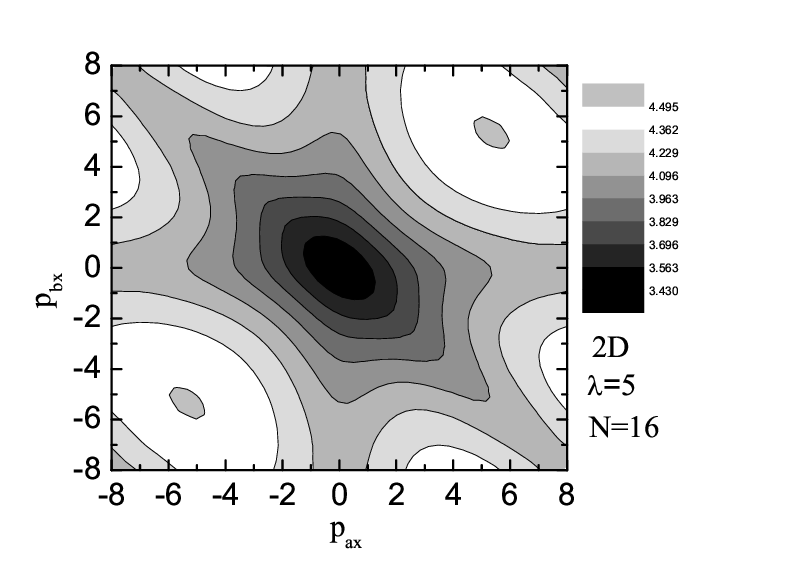}
\includegraphics[scale=0.7]{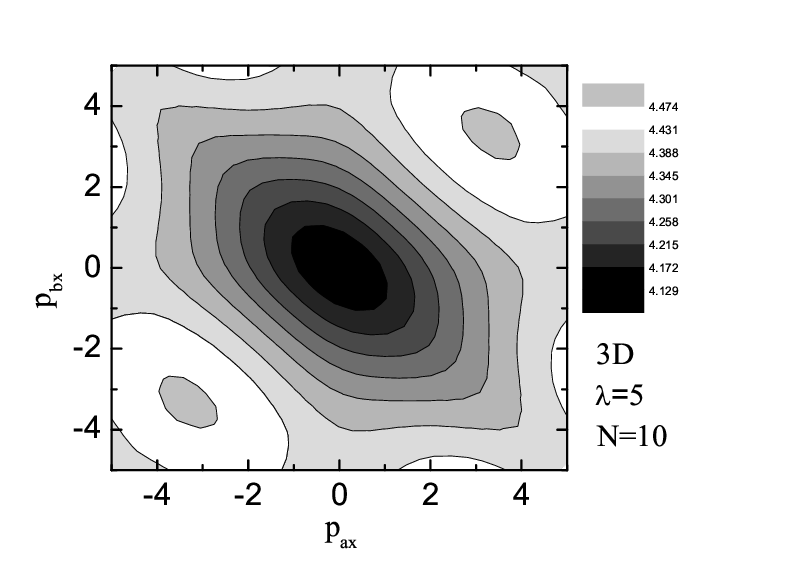}
\caption{Contour plot of the 4-point vertex in 2D and 3D as a
function of $(p_a)_x$ and $(p_b)_x$ with all other momenta
components set to zero for $\lambda=5$.}\label{f9}
\end{figure}

In  figure \ref{f9}  we give contour plots of the 2D and 3D 4-point
vertex which show the dependence of the vertex on the two momentum
components  $(p_a)_x$ and $(p_b)_x$ with all others chosen to be
zero. The vertex has a minimum at the origin of the coordinates, and
the gradient varies with direction.

\section{Summary and Outlook}
\label{concSect}

We have solved the integral equations which determine the
self-energy and vertex functions in 2D and 3D at zero temperature
using a numerical lattice method. 
All results agree
with the perturbative ones when the coupling is small but deviate significantly when the coupling strength increases.
In 2D the 4PI calculations with lattice number $N=16$ are
convergent and the
non-perturbative 4-point vertex and self energy  show similar
asymptotic behaviors at large momentum as the perturbative ones.
 In 3D the 4PI calculations with $N=10$ are reasonably 
well convergent, especially for the 4-point
vertex. To obtain more accurate results at large momenta in 3D we should extend our calculations to larger lattice
number. This requires a different numerical method and increased computing power, and work on this is currently in progress. 

We
comment that zero temperature is the simplest situation numerically,
but not the one in which it is expected that $n$PI methods will have
a substantial advantage over perturbation theory, which is known to
break down at high temperatures. Our calculation makes use of the symmetries of the 2- and 4-point functions, namely the fact that they are symmetric under the interchange of legs, and the interchange of co-ordinate axes. At finite temperature, the number of symmetries will be reduced and the memory requirements will be correspondingly larger.

Our numerical calculations
demonstrate that 4PI calculations are both interesting and feasible,
and motivates further work on more physically interesting problems.

\section*{Acknowledgements}

This work was supported by the Natural and Sciences and Engineering
Research Council of Canada. WJF is supported in part by the National
Natural Science Foundation of China under contract No. 11005138.

\end{document}